\documentclass[10pt,prb,aps,twocolumn,superscriptaddress,floatfix,showpacs]{revtex4}

\pdfoutput=1

\usepackage{graphicx}
\usepackage{amssymb,amsmath}
\usepackage{epstopdf}
\usepackage{bm}
\usepackage{color}

\begin{document}

\title{Correlated valence bond state and its study of the spin-$1/2$
  $J_1-J_2$ Antiferromagnetic Heisenberg model on a square lattice}

\author{Ling Wang} \affiliation{Institute for Quantum Information and
  Matter and Physics Department, California Institute of Technology,
  Pasadena, California 91125, USA}

\date{\today}

\begin{abstract}
  We propose a class of variational wavefunctions, namely the
  correlated valence bond states, for the frustrated Hamiltonians in
  the paramagnetic phase. This class of wavefunctions admits negative
  amplitude and the same sublattice pairing when a bipartite lattice
  is considered, thus suffers from the negative sign problem. However
  if applied to small systems, the sign problem is manageable using
  the standard variational Monte Carlo method. We optimize the
  wavefunctions for the $J_1-J_2$ Antiferromagnetic Heisenberg model
  on a square lattice in the coupling region $J_2/J_1\in [0.45:0.56]$
  for system sizes $L=4,6,8$. To calculate the correlation functions
  and the order parameters for larger systems, we make the extensive
  Monte Carlo samplings using the variational parameters optimized at
  system size $L=8$. We find that the paramagnetic phase is a gapless
  spin liquid in the entire range of $J_2/J_1\in [0.45:0.56]$ with a
  gapless singlet excitation and a gaped triplet excitation.
\end{abstract}

\maketitle

\section{\label{intro}Introduction}
The $J_1-J_2$ antiferromagnetic (AF) Heisenberg model on a square
lattice has attracted a lot of attention due to its close relation to
the disappearance of the AF order in high-T$_c$ superconducting
material~\cite{anderson1,anderson2,Inui,lyons,wen_review_hightc} and
its possibility of realizing the so called spin liquid
state~\cite{chandra88,ed2,ed4,Capriotti_prl87.097201,richter_ed,tao,ling,jiang,huwenjun}. The
spin liquids, defined as lacking long range order and supporting
strong quantum fluctuation, have all the key properties of the
insulating state ``nearby'' to a superconducting phase. Hence a simple
example of spin liquids, the resonating valence bond (RVB) state was
proposed to describe high-T$_c$
superconductivity~\cite{anderson1,anderson2}. Spin liquids have become
the focus of research in modern condensed matter physics since the
discovery of different kinds of spin liquids in
theory~\cite{KL8795,WWZ8913,RS9173,W9164,SondhiZ2,HongZ2}. Although
the existence and stability of spin liquids in realistic models and
real material are still in
question~\cite{kagomeSL,stefan,spinliquid1,spinliquid2,spinliquid3}.

The nature of the ground state of the $J_1-J_2$ AF Heisenberg model in
the intermediate coupling region has been
debated~\cite{gelfand89,chubukov91,chandra88,chandra90,subir,WWZ8913,W9164,ed1,ed2,ed3,ed4,ed5,richter_ed,sirker_prb73.184420,darradi_prb78.214415,kevin_prb79.224431,mambrini_prb74.144422,arlego_prb78.224415,Capriotti_prl87.097201,Capriotti_prl84.3173,valentin,tao,ling}
for decades due to the highly frustrated nature of this model and is
still a open question. Recently, several numerical works revisited
this model using novel concept, such as the topological entanglement
entropy (TEE)~\cite{jiang}, and new numerical tools, such as the
projected entangled pair states
(PEPSs)~\cite{valentin,ling}. Amazingly many of them have reached a
consistent and the lowest ground state energy
ever~\cite{huwenjun,jiang,gong,ling}, however their resulting ground
states behave diversely. A density matrix renormalization group (DMRG)
study on long cylinders suggested a gaped $\mathbb{Z}_2$ spin liquids
from the evidences of non-vanishing singlet and triplet gaps in the
paramagnetic phase and a nonzero TEE~\cite{jiang}. The
Gutzwiller-projected BCS wavefunction study indicated a gapless spin
liquid state supported by the existence of gapless triplet excitations
at momenta $(\pi,0)$ and $(0,\pi)$~\cite{huwenjun}. Most recently, the
DMRG study with $\text{SU}(2)$ symmetry imposed on the states on long
cylinders found a diverging dimer correlation length, and claimed a
plaquette valence bond solid (VBS) state as the ground
state~\cite{gong}. These controversial results pose the question about
the numerical convergence and the system size
dependence~\cite{anders2}.

In this paper, we propose a novel wavefunction and use the variational
Monte Carlo (VMC) method to tackle this system. As one knows, the key
to a meaningful variational method is to have a good trial
wavefunction. Here, we take a resonating valence bond state
approach. The reasons for such a choice are the following, first of
all, RVB states can describe a variety of phases including the spin
liquids, the AF long range ordered states and the valence bond
solids~\cite{KL8795,SondhiZ2,HongZ2,didier1,albuquerque,tang,kevin2,jie1,lin_cap};
second of all, recent development in PEPSs showed that a family of one
parameter PEPSs with bond dimension $D=3$, which described the RVB
state beyond the nearest neighbor pairing, can greatly lower the
variational energy of this model compared to that of the short range
RVB state~\cite{ling_PRL13}. We can imagine that with more build-in
correlations and hence more variational parameters, the RVB states
should approach the true ground state very well.


The simplest RVB ansatz on a bipartite lattice is the valence bond
amplitude product (AP) state, where the probability of having a
valence bond of separation $(x,y)$ is independent of each other and is
denoted as $h(x,y)$~\cite{liang}. Consequently the bond amplitudes
$h(x,y)$ become the variational parameters for the valence bond AP
states. Simulations based on this variational ansatz has been
done~\cite{jie1,kevin2}: it was found that the amplitude $h(1,2)$
tends to be negative for $J_2/J_1\gtrapprox 0.4$ in order to minimize
the energy within this variational subspace. The place where $h(1,2)$
changes sign signals the break down of the Marshall's sign
rule~\cite{marshall} and is very close to the proposed critical point
where the AF long range order disappears. It is interesting to ask the
question that does the onset of the negative amplitude in the valence
bond AP state indicate a phase transition to new states?  In this
paper, we will give our answer to this question from the correlated
valence bond point of view. Prior to our investigation, Lin et
al. introduced the so called correlated amplitude product (CAP) state,
where an extra weight factor has been given to a pair of valence
bonds~\cite{lin_cap} in order to build the bond-bond correlations into
the wavefunction, however they have not yet applied the CAPs to the
$J_1-J_2$ AF Heienberg model. The goal of this paper is to simulate
the ground state of the $J_1-J_2$ AF Heisenberg model on square
lattice using the correlated valence bond states, whose definition
will be clear in Sec~\ref{wf}.

The rest of this paper is arranged as following: in Sec.~\ref{wf}, we
define the correlated valence bond states, which have some similarity
but are essentially different from the CAPs~\cite{lin_cap}. We
introduce the variational Monte Carlo method and the optimization
strategy for this wavefunction in Sec.~\ref{methods}. Upon
optimization of the correlated valence bond states by minimizing the
ground state energies, we arrive at the optimal states. In
Sec.~\ref{results}, We present the variational ground state energy and
the ground state correlations for the optimized states. In
Sec.~\ref{discussion}, we discuss the advantages and limitation of
this ansatz and conclude for the nature of the intermediate phase of
the $J_1-J_2$ AF Heisenberg model on square lattice.

\section{\label{wf}Hamiltonian and the trial wavefunction}
The Hamiltonian of the $J_1-J_2$ AF Heisenberg model on a square
lattice is given by
\begin{equation}
\label{afheisenberg}
  H=J_1\sum_{\langle i,j\rangle}\mathbf{S}_i\cdot\mathbf{S}_j+J_2\sum_{\langle\langle i,j\rangle\rangle}\mathbf{S}_i\cdot\mathbf{S}_j,
\end{equation}
where the first summation runs over the NN pair $\langle i,j\rangle$
and the second summation runs over the next NN (NNN) pair
$\langle\langle i,j\rangle\rangle$. 

\begin{figure}
\begin{center}
\includegraphics[width=\columnwidth]{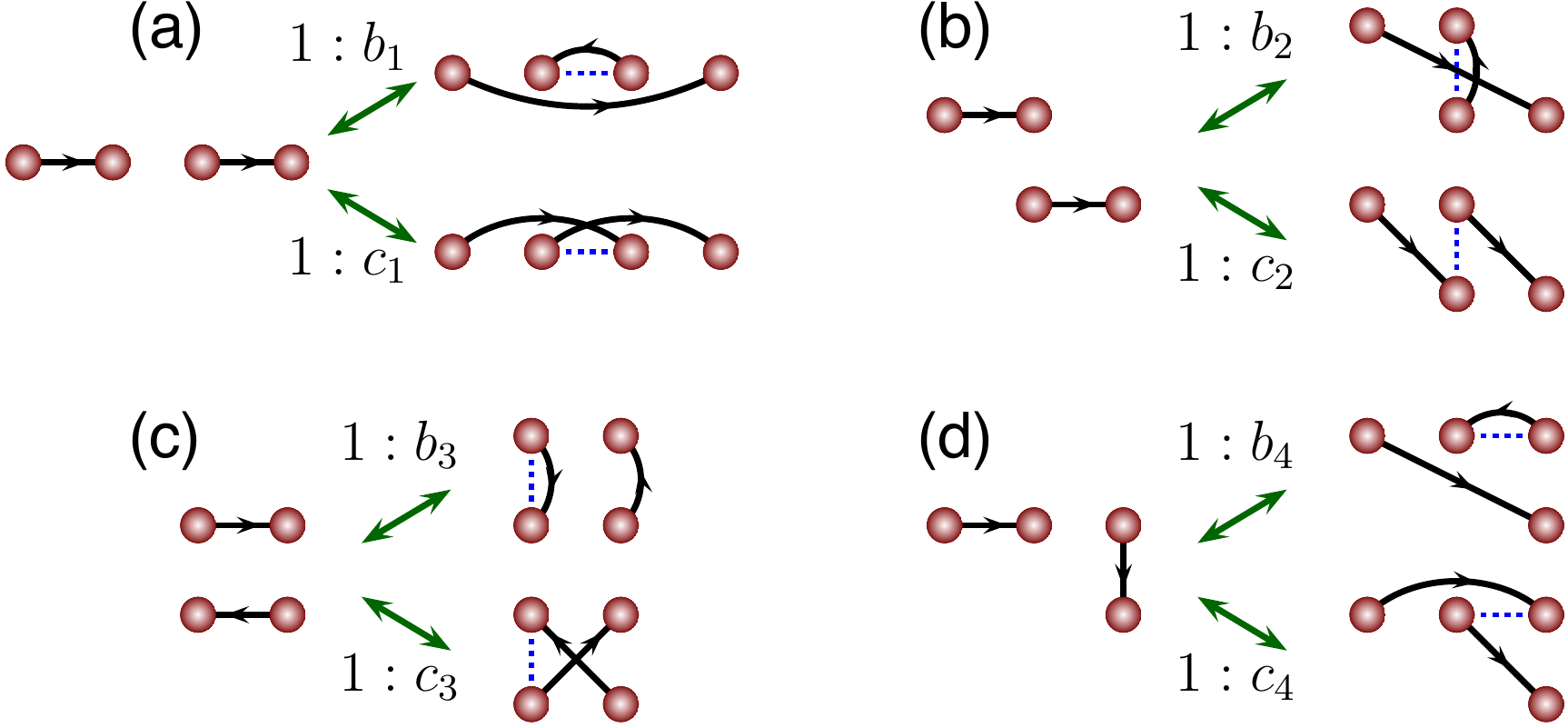}
\caption{ On the left hand side of the reversible arrows, we
  demonstrate four possible short bond pairs which are connected by a
  common edge in the lattice; whereas one the right hand side of the
  arrows, we show two other valence bond tilings of the four sites.
  Blue dashed lines are the reference edge. The ratios mark the
  relative amplitudes of the short bonds and the correlated-bond-pairs
  in the wavefunction coefficients.}
\label{cap}
\end{center}
\end{figure}

An equal weight superposition of the short range RVB state has been
applied as a trial wavefunction to the ground state of Hamiltonian
Eq.~(\ref{afheisenberg})~\cite{didier1}, but the variational energy is
not good and it is nowhere close to the true ground state at any
coupling ratio $J_2/J_1$. One step forward was made by introducing a
positive amplitude to an arbitrary ranged bipartite valence bond in
the frame work of the valence bond AP states~\cite{kevin2}; the
reported best thermodynamic energy at $J_2/J_1=0.5$ is
$E_{\text{AP}}=-0.49023(2)$ per site, which is still much higher than
the DMRG studies~\cite{gong} and has left a lot of room for
improvement in the trial wavefunction. Here we speculate that the
bond-bond correlations may play a crucial role than the individual
long range bonds~\cite{lin_cap}. Therefore, we define a class of
ansatze called the correlated valence bond states by a set of
statistical rules:
\begin{enumerate}
\item Considering without the presence of the correlated valence bond
  pairs, the ansatz is reduced to a short-bond amplitude product
  state: the bond amplitude $h(1,0)=h(0,1)=1$ and otherwise 0.

\item If two short individual bonds are sitting next to a common edge
  of the lattice, they are allowed to quantum fluctuate to any other
  valence bond tilings on these four sites, and {\it vice versa}, as
  illustrated in Fig.~\ref{cap}. Each valence bond tiling of these
  four sites is associated with a amplitude marked as either 1, $b_i$
  or $c_i$, where $b_i,c_i$ ($i=1,2,3,4$) are the variational
  parameters.

\item The two short bonds on the left hand side of the
  reversible-arrow in Fig.~\ref{cap} are two individual bonds
  (uncorrelated), which can be randomly rearranged with all other
  uncorrelated short bonds into any allowed short bond product state.
  However the two correlated bonds together with the lattice edge (in
  blue dash line) on the right hand side of the reversible-arrow form
  a single object such that the two bonds there in can not be treated
  independently. The correlated-bond-pair can only be flipped back to
  the two short bonds exactly shown on the left hand side of the
  reversible-arrow.
\end{enumerate}
The above three rules together form a statistical mechanic
ensemble. Before discussing the variational Monte Carlo algorithm,
several comments are along the line. We choose such a statistical
ensemble as the trial wavefunction in order to address bond-bond
correlations. The amplitudes of the correlated-bond-pairs give extra
factors to the wavefunction coefficients, and this ansatz goes beyond
the AP state. The correlated-bond-pair amplitudes serve as the
variational parameters of the trial wavefunction. We can easily
translate this ansatz to a PEPS wavefunction that describes
qualitatively the same statistical ensemble.

We impose the translational and rotational symmetries to the
wavefunction to reduce the number of variational parameters. This
restriction is perfectly valid as long as we work with finite size
systems. There are four distinct arrangements of two short bonds that
are connected by a common lattice edge, as illuminated in
Fig.~\ref{cap}, therefore eight different correlated-bond-pairs whose
amplitudes are denoted by $b_1,b_2,b_3,b_4$ and
$c_1,c_2,c_3,c_4$. Hereafter, we use the word ``correlated-bond-pair''
and its amplitude interchangeably. Given these clarifications, we can
write the trial wavefunction as following
\begin{equation}
\label{crvb}
|\Psi\rangle=\sum_{C(\alpha)}\prod_{i=1}^4b_i^{n_{b_i}}\prod_{i=1}^4c_i^{n_{c_i}}|\alpha\rangle\equiv\sum_{C(\alpha)}\psi_{C(\alpha)}|\alpha\rangle,
\end{equation}
where $C(\alpha)$ is a compact pack of individual short bonds and
correlated-bond-pairs on a lattice which produce a valence bond tiling
configuration $|\alpha\rangle$, $n_{b_i}$ ($n_{c_i}$) is the number of
$b_i$ ($c_i$) in a compact pack $C(\alpha)$. Note that different
packing $C(\alpha)$ can generate the same valence bond tiling
configuration $|\alpha\rangle$.

\begin{figure}
\begin{center}
\includegraphics[width=0.85\columnwidth]{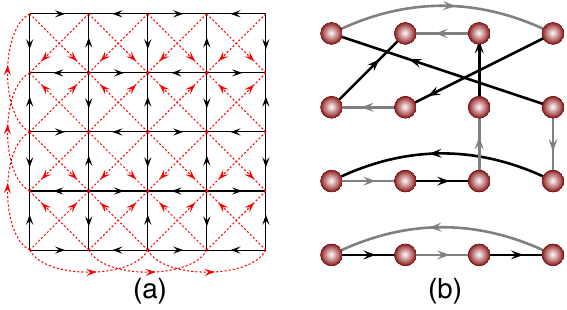}
\caption{(a) The singlet sign convention for the wavefunction. (b) A
  transition graph for calculation of the overlap matrix element
  $\langle {\beta}|{\alpha}\rangle$. The arrows marked for
  $|\alpha\rangle$ (in black lines) align with the sign convention,
  whereas for $|\beta\rangle$ (in gray lines) anti-align with the sign
  convention. $\langle
  {\beta}|{\alpha}\rangle=(-1)^{n_v}2^{n_l-\frac{N}{2}}$, where $n_v$
  is the number of arrows that violate the direction of the flow and
  $n_l$ is the number of loops in the transition graph.}
\label{sign}
\end{center}
\end{figure}

\section{\label{methods}The variational Monte Carlo method}
In this section, we first briefly describe the variational valence
bond Monte Carlo method, followed by the optimization method used to
determine the optimal variational parameters.

Given a variational wavefunction $|\Psi\rangle$, the expectation value
of any operator $\hat{O}$ is written as
\begin{equation}
\label{exp}
\frac{\langle\Psi|\hat{O}|\Psi\rangle}{\langle\Psi|\Psi\rangle}=\frac{\sum_{C(\alpha)C(\beta)}W_{C(\alpha)C(\beta)}O_{\alpha\beta}}{\sum_{C(\alpha)C(\beta)}W_{C(\alpha)C(\beta)}},
\end{equation}
with the importance sampling weight defined as
\begin{equation}
\label{wt}
W_{C(\alpha)C(\beta)}=\psi_{C(\beta)}\psi_{C(\alpha)}\langle \beta |\alpha\rangle,
\end{equation}
and the operator matrix element defined as
\begin{equation}
O_{\alpha\beta}=\frac{\langle \beta|\hat{O}|\alpha\rangle}{\langle \beta|\alpha\rangle},
\end{equation}
here $\langle \beta|\alpha\rangle$ is the overlap matrix element. If
both the coefficients $\psi_{C(\alpha)}$ and the overlap matrix
elements $\langle\beta|\alpha\rangle$ are positive definite, we could
use the standard valence bond MC method~\cite{jie1}. However our
valence bond configuration $|\alpha\rangle$ allows valence bonds of
the same sublattice pairing, therefore the overlap matrix elements can
be negative; in addition, the correlated-bond-pair amplitudes can take
negative values, which means the coefficients $\psi_{C(\alpha)}$ can
be negative too. Thus our wavefunction encounters a negative sign
problem. We can treat it in a standard way by rewriting
Eq.~(\ref{exp}) as
\begin{eqnarray}
\label{exp2}
\nonumber
\frac{\langle\Psi|\hat{O}|\Psi\rangle}{\langle\Psi|\Psi\rangle}&=&\frac{\frac{\sum_{C(\alpha)C(\beta)}|W_{C(\alpha)C(\beta)}|\text{Sgn}({W})O_{\alpha\beta}}{\sum_{C(\alpha)C(\beta)}|W_{C(\alpha)C(\beta)}|}}{\frac{\sum_{C(\alpha)C(\beta)}|W_{C(\alpha)C(\beta)}|\text{Sgn}({W})}{\sum_{C(\alpha)C(\beta)}|W_{C(\alpha)C(\beta)}|}}\\
&\equiv&\frac{\langle\text{Sgn}(W)O_{\alpha\beta}\rangle_{|W|}}{\langle\text{Sgn}(W)\rangle_{|W|}},
\end{eqnarray}
where $\text{Sgn}(W)$ denotes the sign of the weight Eq.~(\ref{wt}).
We sample using the absolute value $|W_{C(\alpha)C(\beta)}|$
(abbreviated as $|W|$) as the importance weight.  The operator
expectation value can be obtained by taking the ratio of
$\langle\text{Sgn}(W)O_{\alpha\beta}\rangle_{|W|}$ and
$\langle\text{Sgn}(W)\rangle_{|W|}$.

We now explain how to calculate the overlap matrix element
$\langle\beta|\alpha\rangle$ and how to count the sign in
$\text{Sgn}(W)$. We fix the sign convention for any valence bond in
$|\alpha\rangle$: for the AB sublattice pairing, the direction of the
valence bond is always pointing from the A sublattice to the B
sublattice; whereas for the same sublattice pairing, we refer to the
signs drawn in Fig.~\ref{sign}(a) as the convention. Given
$|\alpha\rangle$ and $|\beta\rangle$, the overlap matrix element for a
transition graph is $(-1)^{n_v}2^{n_l-\frac{N}{2}}$, where $n_l$ is
the number of loops in the transition graph and $n_v$ is the number of
singlets that violate the direction of flow that is arbitrarily chosen
for each loop in the transition graph. An example of the transition
graph is shown in Fig.~\ref{sign}(b). Here $2^{-N/2}$ is a
normalization constant, since the maximum number of loops in a
transition graph is $N/2$. The total number of minus sign factors in
the coefficient $\psi_{C(\alpha)}$ depends on the sign convention and
the signs of the variational parameters $b_i,c_i$. Putting all signs
together we have
$\text{Sgn(W)}=\text{Sgn}(\psi_{\alpha})\text{Sgn}(\psi_{\beta})\text{Sgn}(\langle\beta|\alpha\rangle)$.

Next, we explain how to calculate the operator matrix element
$O_{\alpha\beta}$. Since the singlet product states
$|\alpha\rangle,|\beta\rangle$ appear on both the numerator and the
denominator, we can choose any singlet sign convention we want (those
sign factors will cancel if otherwise choose a difference sign
convention). For any transition graph, we choose one such that every
loop has $A^{\prime}B^{\prime}A^{\prime}B^{\prime}\cdots
A^{\prime}B^{\prime}$ structure, {\it i.e.} we choose the sign to be
always pointing from $A^{\prime}$ to $B^{\prime}$. Here we use
$A^{\prime},B^{\prime}$ to differentiate the $A,B$ sublattices of the
original lattice bipartition, and $A^{\prime}$ and $B^{\prime}$ only
have a relative meaning within each loop. Therefore all operator
expectation values can be evaluated using the same formulas as in the
Ref.~\cite{anders2} with the replacement of $A$ ($B$) by $A^{\prime}$
($B^{\prime}$).

\begin{figure}
\begin{center}
\includegraphics[width=\columnwidth]{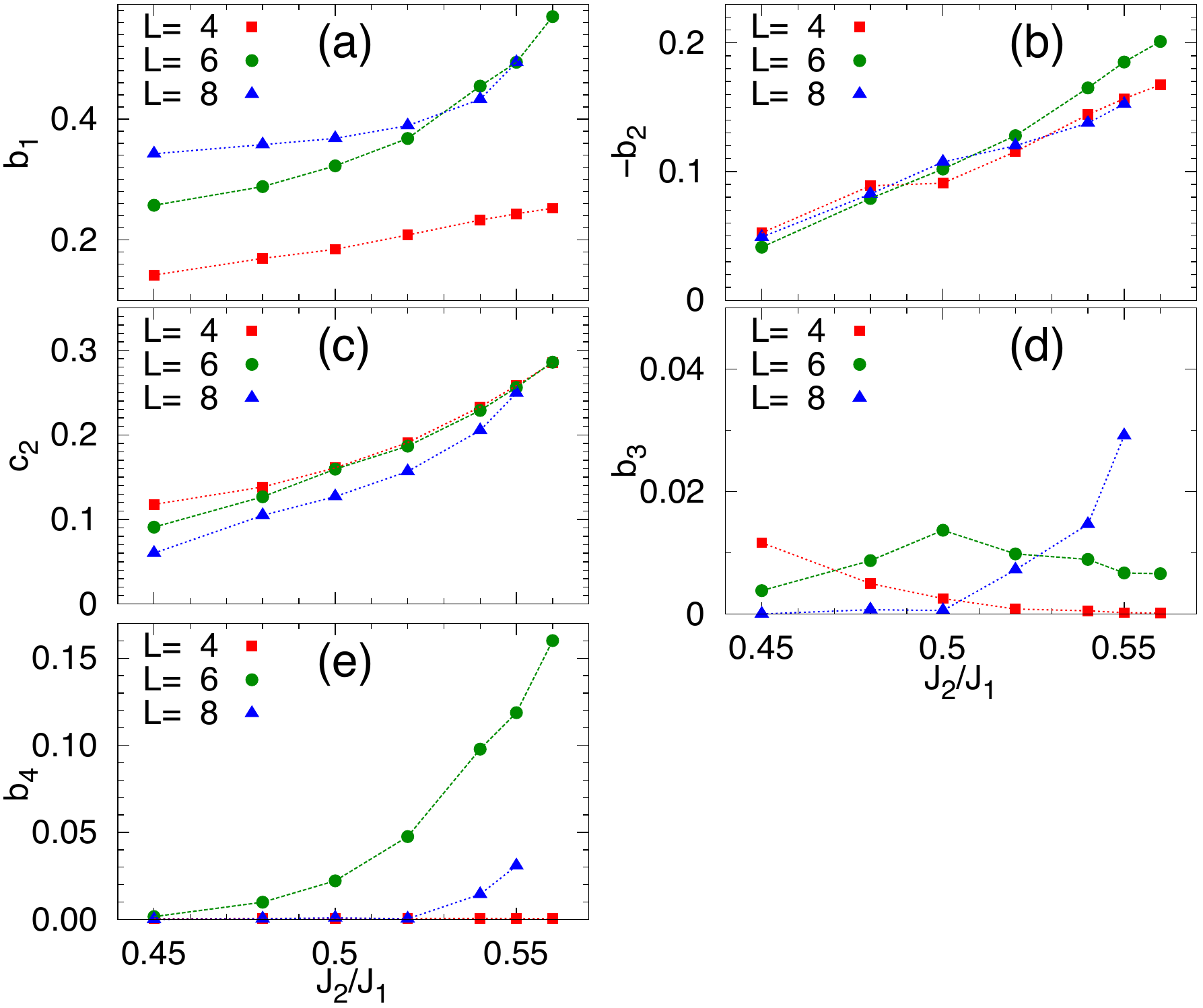}
\caption{Optimized variational parameters for system sizes $L=4,6,8$
  with coupling strength $J_2/J_1\in [0.45:0.56]$. Parmeters not shown
  here are optimized close to 0.}
\label{para}
\end{center}
\end{figure}

The MC sampling of wavefunction Eq.~(\ref{crvb}) takes two types of
update. The first type of update is the local update. We begin by
randomly selecting a link from the lattice, one of the following cases will
happen:
\begin{enumerate}
\item if the link is not occupied by any singlets and it sits beside
  two individual short bonds, propose to flip randomly to one of the
  correlated-bond-pairs or stay unchanged with the corresponding
  probability;
\item if the link belongs to a correlated-bond-pair and is exactly
  sitting on the reference edge in blue dashed lines in
  Fig.~\ref{cap}, propose to flip to the two individual short bonds or
  stay unchanged with the corresponding probability;
\item for any other cases, abandon such a choice and iterate.
\end{enumerate}
If a local update is proposed, we accept or reject this local update
with probability $P=\text{min}[2^{n_l^{\prime}-n_l},1]$, where
$n_l^{\prime}$ is the number of loop in the trial transition
graph. The second type of update is the loop update. We randomly
construct an allowed loop of alternating individual short bonds and
vacant lattice links, then we shift all individual short bonds on that
loop by one lattice space. This rearrangement can always be proposed
with probability 1, since all individual short bond have equal
amplitudes, and it will keep all the correlated-bond-pairs
untouched. The loop update is accepted or rejected with probability
$P=\text{min}[2^{n_l^{\prime}-n_l},1]$ defined above.

We use variational MC to calculate the derivative of the energy with
respect to a variational parameter $a$ ($a\in\{b_i,c_i\}$) as
\begin{equation}
\frac{\partial \langle E \rangle}{\partial a}=\langle\frac{n_a}{a}E\rangle-\langle\frac{n_a}{a}\rangle\langle E\rangle,
\end{equation}
and update them according to~\cite{jie1}
\begin{equation}
a({t+1})=a({t})-r\delta({t+1})\times \text{sign}\left(\left(\frac{\partial \langle E\rangle}{\partial a}\right)_t\right),
\label{randomupdate}
\end{equation}
where $t$ is the iteration index, $r$ is a random number $r\in[0,1)$,
and $\delta({t})=0.01/t$ until the energy converges. The function form
$\delta(t)$ is chosen heuristically~\cite{jie1}. The optimization
results and correlation functions will be presented in the next
section.

\begin{figure}
\begin{center}
\includegraphics[width=.9\columnwidth]{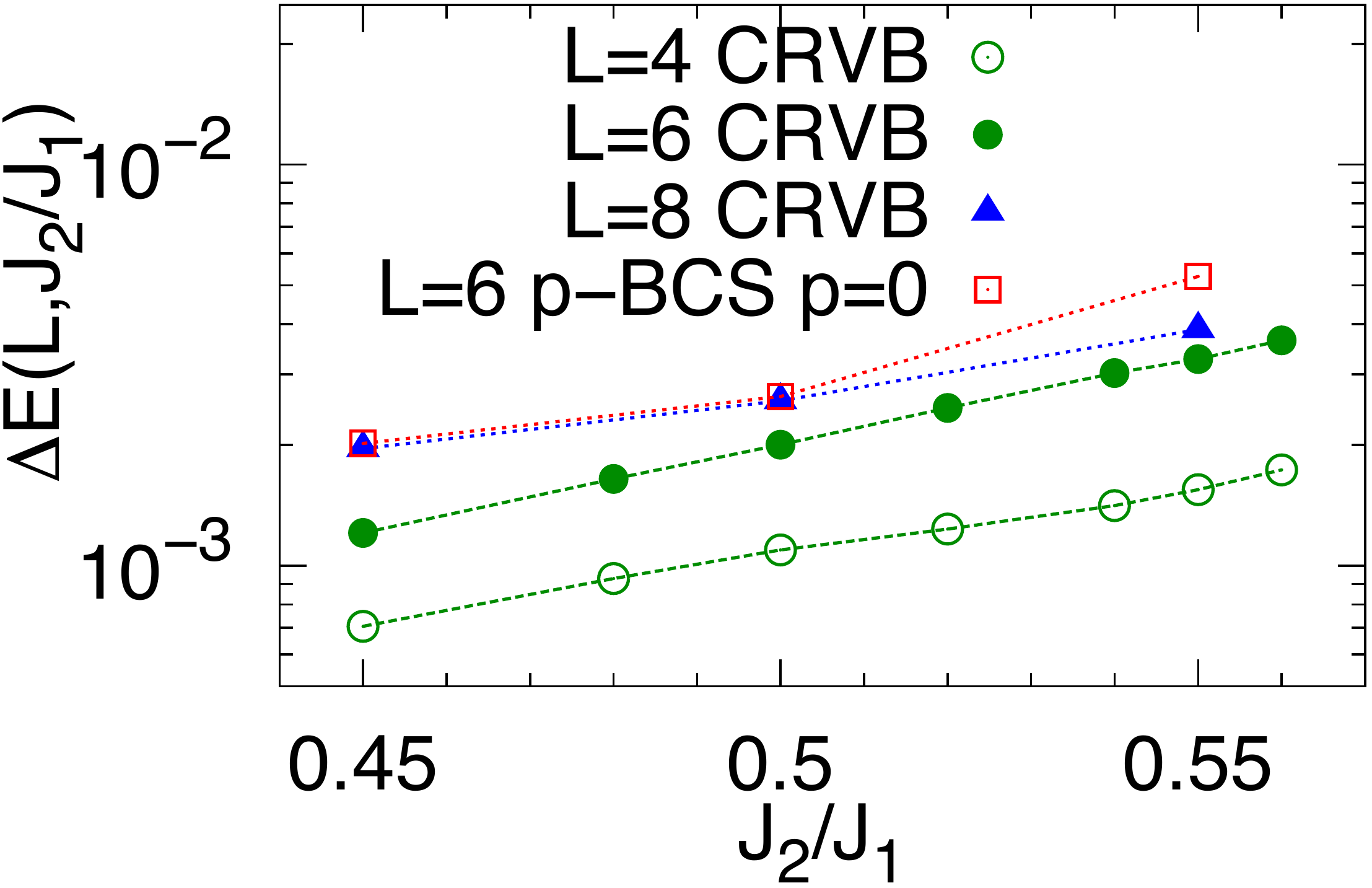}
\caption{The ground state energy error per site for the $4\times 4$
  (green circle), $6\times 6$ (green dots) and $8\times 8$ (blue
  triangles) lattices, here the reference ground state energies for
  the $8\times 8$ lattice are taken from the DMRG results in
  Ref.~\cite{gong}. Red square is a comparison to the projected
  fermionic BCS wavefunction for a $6\times 6$ lattice without Lanczos
  projection step~\cite{huwenjun}.}
\label{err}
\end{center}
\end{figure} 

\section{\label{results}The variational Results} 
We present the well optimized variational parameters for the
wavefunction Eq.~(\ref{crvb}) for system sizes $L=4,6,8$ at
$J_2/J_1\in[0.45:0.56]$ in Fig.~\ref{para}. We find strong size
dependence for the optimized variational parameters
$b_1,b_3,b_4$. However, parameters $b_2,c_2$ have less size
dependence. Other parameters that are not shown in Fig.~\ref{para} are
optimized close to zero. The absolute energy error per site $\Delta
E(L,J_2)$ for sizes $L=4,6$ compared with the exact diagonalization
(ED) results~\cite{richter_ed} and for size $L=8$ compared with the
DMRG results~\cite{gong} are presented in Fig.~\ref{err}. The absolute
errors are small, {\it e.g.}, $\Delta E(4,0.55J_1)\sim 1.5\times
10^{-3}J_1$ and $\Delta E(6,0.55J_1)\sim 3\times 10^{-3}J_1$. For
comparison, we draw the Gutzwiller-projected BCS wavefunction results
without Lanczos projection~\cite{huwenjun} in the same frame: at
$L=6,J_2=0.55J_1$, our absolute energy error is lower by about
$40\%$. 

We take the optimized parameters from size $L=8$ to calculate the
order parameters and the correlation functions for system sizes $L>8$
until the average signs are no longer manageable, because optimizing
variational parameters for $L>8$ becomes not feasible. We assume that
the correlation functions at larger sizes will not be too sensitive to
the variational parameters as long as they are within a reasonable
range from the optimal values. 

\begin{figure}
\begin{center}
\includegraphics[width=.9\columnwidth]{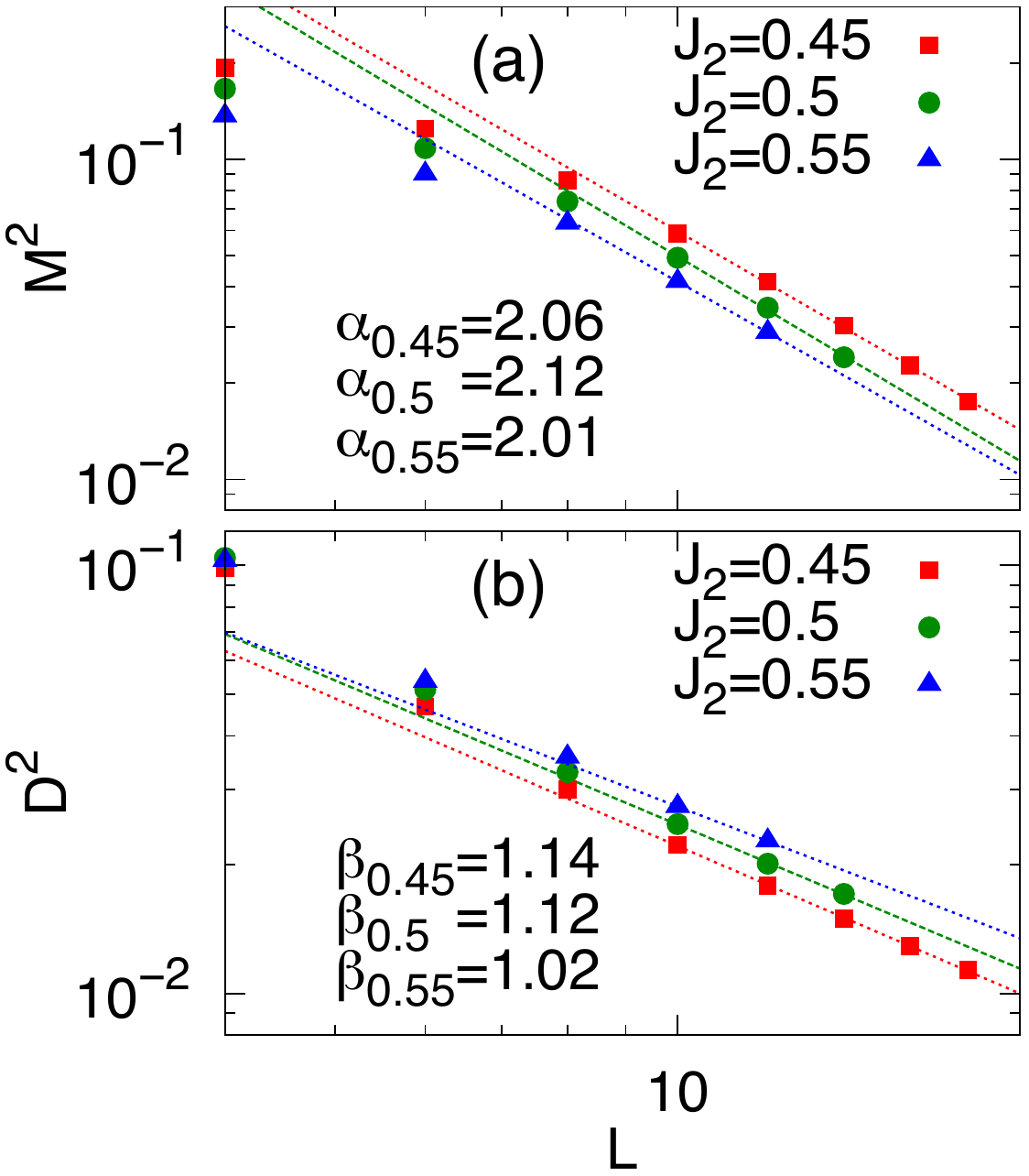}
\caption{(a) The sublattice magnetization and (b) the dimer order
  parameter as a function of system size at $J_2/J_1=0.45,0.5,0.55$ in a
  log-log plot. Power law functions $M^2\sim 1/L^{\alpha}$, $D^2\sim
  1/L^{\beta}$ have been fitted to the points for $L\geq 10$.}
\label{ordpara}
\end{center}
\end{figure}

Let us first define a set of order parameters. The sublattice
magnetization is written as
\begin{equation}
M^2=\frac{1}{N}\sum_{\mathbf{r}}C(\mathbf{r}),
\end{equation}
where $\mathbf{r}=(x,y)$, $C(\mathbf{r})$ is the spin correlation
function, and it is defined as
\begin{equation}
\label{ssc}
C(\mathbf{r})=\frac{(-1)^{x+y}}{N}\sum_{i=1}^N\mathbf{S}_{\mathbf{r}_i}\cdot\mathbf{S}_{\mathbf{r}_i+\mathbf{r}}.
\end{equation}
The dimer order parameter is defined as
\begin{equation}
D^2=D_x^2+D_y^2,
\end{equation}
where
\begin{eqnarray}
D_x&=&\frac{1}{N}\sum_i^N(-1)^{x_i}\mathbf{S}_{\mathbf{r}_i}\cdot\mathbf{S}_{\mathbf{r}_i+\mathbf{e}_x},\\
D_y&=&\frac{1}{N}\sum_i^N(-1)^{y_i}\mathbf{S}_{\mathbf{r}_i}\cdot\mathbf{S}_{\mathbf{r}_i+\mathbf{e}_y},
\end{eqnarray}
and $\mathbf{e}_x=(1,0)$, $\mathbf{e}_y=(0,1)$.

The sublattice magnetization and the dimer order parameter are
presented in log-log plots in Fig.~\ref{ordpara}. The sublattice
magnetization at couplings $J_2/J_1=0.45,0.5,0.55$ scale as $M^2\sim
1/L^2$, which is expected for the exponentially decaying spin
correlation function Eq.~(\ref{ssc}). The dimer order parameter
follows a power law decay with system size as $D^2\sim 1/L^{\beta}$
where $\beta\approx 1$. Therefore, we find a critical phase that has
gapless singlet excitations and gaped triplet excitations within the
range of $J_2/J_1\in [0.45:0.56]$.

We next make some connections to the results given by previous work. A
large optimal parameter $b_1$ is consistent with the results of an
enhanced spin-spin correlation along the $x$ and $y$ axes shown in
Ref.~\cite{kevin2}. A negative optimal parameter $b_2$ is consistent
with the result of having negative $h(1,2)$ in the simulation using an
AP product state~\cite{jie1,kevin2}. Our variational term $c_2$
generates configuration of parallel diagonal bond pair. The parallel
diagonal bond pairs together with the short bonds, if contribute with
equal weights to the resonating wavefunction on a square lattice, will
give a $\mathbb{Z}_2$ spin liquid state~\cite{HongZ2}. Our optimized
wavefunction, which contains both the parallel diagonal bonds and the
short bonds, does not behave like a $\mathbb{Z}_2$ spin liquid,
although it deviates from the special point of equal weights
superposition defined in Ref.~\cite{HongZ2}. We need to do further
investigation to answer why it fails to be a $\mathbb{Z}_2$ spin
liquid. The optimal parameter $b_3$ at system size $L=8$ grows rapidly
as a function of $J_2$, whose effect is to increase the dimer-dimer
correlation. However given this optimized parameter strength, the
effect of $b_3$ will not induce a quantum phase transition from a
critical phase to the VBS phase as predicted in Ref~\cite{gong},
because the latter phase requires a very large value in
$b_3$~\cite{lin_cap}.

Let us turn to the question that we asked earlier: does a negative
amplitude $h(2,1)$ in an AP state indicate a phase transition?  From
our example, the answer seems to be NO. The critical phase presented
in our work is simply a result of muting all the long range bonds in
the AP state, as one increase the weights of $b_2$ and $c_2$ from 0,
there are no signs of a phase transition from the critical phase to
either $\mathbb{Z}_2$ or VBS states. Therefore we conclude that the
negative amplitude $h(2,1)$ along in an AP state could not trigger a
phase transition.

\section{\label{discussion}Conclusion}
Using the correlated valence bond state, we minimized the ground state
energy of the $J_1-J_2$ antiferromagnetic (AF) Heisenberg model on a
square lattice for a coupling ratio $J_2/J_1\in [0.45:0.56]$ by
turning a few correlated-bond-pair amplitudes. The energies are
consistent with the exact diagonalization (ED) results on the $4\times
4$ and $6\times 6$ tori~\cite{richter_ed} and the density matrix
renormalization group (DMRG) results on the $8\times 8$
torus~\cite{gong}. We applied the optimal variational parameters from
the $8\times 8$ system to the larger tori and studied their
correlation functions using the valence bond Monte Carlo (MC) sampling
method. We found that within the optimized phase, the Neel order
parameter scales as the inverse of the volume, and the dimer order
parameter describing the columnar or plaquette valence bond solid
(VBS) phases follows a power law decay with the system size $L$
approximately as $1/L$. These correlation functions indicate a
critical phase with gapless singlet excitations and gaped triplet
excitations in the entire range of $J_2/J_1\in [0.45:0.56]$. Due to
the negative sign problem, we can not optimize even larger systems or
further increase the number of variational parameters, such as turn on
the individual long range bonds. Our simulation provide insights of
how and when the critical phase can turn into a VBS or a
$\mathbb{Z}_2$ spin liquid. However with the current results, we can
not conclude which one is the true ground state with the intermediate
coupling strength.

{\it Acknowledgment}-- We thank J. Richter for providing us the ED
results. We would like to thank O. Motrunich, F. Verstraete, K. Beach,
A. Sandvik, Z.-C. Gu, S.-S. Gong, W.-J. Hu for useful discussion. The
author especially thanks O. Motrunich for reading and giving comments
on the manuscript. This work was supported by the Institute for
Quantum Information and Matter, an NSF Physics Frontiers Center with
support of the Gordon and Betty Moore Foundation through Grant
GBMF1250.

\end{document}